\documentclass[12pt]{iopart}

\usepackage{iopams}
\usepackage{braket}
\usepackage{amsfonts}
\usepackage{color}
\usepackage[dvipsnames]{xcolor}
\usepackage{amssymb}
\usepackage{mathrsfs}
\usepackage{graphicx}
\usepackage{lmodern}
\usepackage{lineno}
\usepackage{hyperref}
\usepackage[normalem]{ulem}

\newcommand{\beq}{\begin{eqnarray}}
\newcommand{\eeq}{\end{eqnarray}}

\newcommand{\nn}{\nonumber}

\def\keywords#1{\vspace{10pt}
     \begin{indented}
     \item[]\rm Keywords: #1\par
     \end{indented}}

\begin{document}



\title{Polymer Quantum Mechanics as a Deformation Quantization}
\author{Jasel Berra--Montiel$^{1,2}$ and 
Alberto Molgado$^{1,2}$}

\address{$^{1}$ Facultad de Ciencias, Universidad Aut\'onoma de San Luis 
Potos\'{\i} \\
Av.~Salvador Nava S/N Zona Universitaria, San 
Luis Potos\'{\i}, SLP, 78290 Mexico}
\address{$^2$ Dual CP Institute of High Energy Physics, Mexico}

\eads{\mailto{\textcolor{blue}{jasel.berra@uaslp.mx}},\ 
\mailto{\textcolor{blue}{alberto.molgado@uaslp.mx}}\ 
}






\begin{abstract}
We analyze the polymer representation of quantum mechanics within the deformation quantization formalism.  In particular, we construct the Wigner function  and the star-product for the polymer
representation as a distributional limit of the 
Schr\"odinger representation for the Weyl 
algebra in a Gaussian weighted measure, and  
we observe that the quasi-probability distribution limit
of this Schr\"odinger representation agrees with the Wigner function for Loop Quantum Cosmology.  Further, 
the introduced polymer star-product fulfills   Bohr's correspondence principle even though not all the operators are well defined in the polymer representation. 
Finally, within our framework, we also derive a generalized uncertainty principle which resembles the one appearing in different scenarios, including theories with a minimal length.
\end{abstract}


\maketitle

\section{Introduction}

The polymer representation of quantum mechanics is obtained by applying non-regular representation techniques developed within the Loop Quantum Gravity (LQG) framework to systems with a finite number of degrees of freedom~\cite{shadow},~\cite{CTZ}. Being a completely background independent quantization scheme, the LQG approach enforce the diffeomorphism covariance at the quantum mechanical level through the unitary implementation of diffeomorphisms on a Hilbert space. The existence of diffeomorphism invariant states suggests that a non-regular representation of the canonical commutation relations must be considered in order to obtain a covariant representation of quantum mechanics. This non-regularity condition, applied to minisuperspace models, has resulted in a polymer-type representation known as Loop Quantum Cosmology (LQC). Under this approach, important advances in the quantum gravity program have been achieved, including the classical singularity avoidance by a quantum bounce~\cite{Bojowald},~\cite{Bojowald1}, a microscopic basis for the black hole entropy~\cite{Rovelli},~\cite{ABC},~\cite{Domagala}, and inhomogeneus perturbations in the cosmic expansion~\cite{Agullo} (and references therein). Despite such efforts
at the quantum level, a proper semiclassical limit of a quantum theory of gravity remains an open issue. The main reason lies in the difficulty to construct semiclassical states in the kinematical Hilbert space  which 
are peaked on classical solutions, and such that they solve the classical dynamics. 
For cosmological models, these difficulties are tackled by performing the quantization in minisuperspace, this means after reducing the phase space and fixing all the kinematical symmetries through the homogeneity conditions for Bianchi models~\cite{Ashtekar}.  Nevertheless, the interplay between these non-regular descriptions and the ordinary Schr\"{o}dinger representation, as well as the correspondence from quantum to classical algebraic structures still need to be deciphered. Only a complete understanding of these issues will allow us to construct a fully background independent quantum dynamics for systems such as general relativity.  

In order to shed some light on  the issues discussed above, in this work we analyze the polymer representation of quantum mechanics within the deformation quantization formalism. The  deformation quantization program, also referred to as  phase space quantum mechanics by many authors, consists in a formal passage from classical to quantum systems using the Dirac quantization framework as a fundamental guideline \cite{BF1}, \cite{BF2}. The idea behind this formalism lies in a  deformation with respect to some parameter (e.g. the Planck constant $\hbar$) of the algebraic and geometrical structures of the classical phase space. Since, for any classical system these structures can be defined in terms of the algebra of observables, that is, smooth real or complex valued functions on the phase space, a deformation  is characterized by a non-commutative product, denoted as the star-product, 
which substitutes the standard point-wise product and contains the necessary quantum information of 
a given system. Consequently, this deformed 
star-product induces 
a deformation of the Poisson bracket  in such a manner that it contains all the information related with the commutator between self-adjoint operators~\cite{Blaszak}. One crucial element of this formulation of quantum mechanics resides on the definition of the Wigner function, which is a quasi-probability distribution function in phase space and corresponds to a representation of the density matrix that is responsible of all the auto-correlation functions of a given quantum mechanical system. As it is well known, within the deformation quantization formalism the language of operators and wave functions, common in quantum mechanics, is interchanged by 
 the $\ast$-product and the Wigner function that  encode the entire quantum mechanical properties, that is, expectation values of observables and transition amplitudes.  This interchange between the classical and quantum structures of a system is realized in practice by the Wigner-Weyl map which is a homomorphism relating classical observables to quantum self-adjoint operators~\cite{Curtright}.

Following some ideas developed in \cite{CTZ}, the aim of this paper is                   
to obtain the Wigner function and the  star-product of the polymer representation as a distributional limit of the Schr\"{o}dinger representation. As we will demonstrate below,
this quasi-probability distribution limit agrees with the Wigner function for LQC, constructed by means of cylindrical functions defined on the Bohr compactification on the real line~\cite{Sahlmann},\cite{Perlov}.  Following the quantization deformation scheme we construct a polymer star product, which in the classical limit $\hbar\mapsto 0$ reduces to  the standard Poisson bracket for smooth functions, thus fulfilling  Bohr's correspondence principle. Finally, we
also derive the uncertainty principle, which happens to be  related to the  Generalized Uncertainty Principles (GUP) that one may encounter in theories based on the existence of a fundamental minimal length~\cite{Kempf}, \cite{Camelia}.     

The paper is organized as follows, in Section 2 we introduce the formalism of deformation quantization focusing our attention in the case of Gaussian measures. In Section 3, we derive the polymer representation as a limiting case of the Sch\"{o}dinger representation within the deformation quantization formalism. In Section 4, the uncertainty principle for the polymer representation is presented. Finally, we introduce some concluding remarks in Section 5.

\section{Deformation quantization in the Gaussian measure}
\label{sec:WWquantization}

In this section, we derive the Wigner-Weyl quantization scheme in the Gaussian measure. For simplicity we restrict our attention to systems with one degree of freedom, but the generalization to more dimensions follows straightforwardly.

\subsection{The Weyl transform}

 The simplest approach to quantization of a classical system, is to provide a one-to-one mapping $Q_{\hbar}:\mathcal{A}\rightarrow \mathfrak{A}$ from the set of classical observables $\mathcal{A}=C^{\infty}(\mathbb{R}^{2})$, to the set of quantum observables $\mathfrak{A}$, given by self-adjoint operators defined on a Hilbert space $\mathcal{H}$. The map $Q_{\hbar}$ must satisfy the properties
\begin{equation}
\lim_{\hbar\to0}\frac{1}{2}Q_{\hbar}^{-1}\left( Q_{\hbar}(f_{1})Q_{\hbar}(f_{2}) + Q_{\hbar}(f_{2})Q_{\hbar}(f_{1})\right)=f_{1}f_{2} \,, 
\end{equation}
and
\begin{equation}
\lim_{\hbar\to 0}Q_{\hbar}^{-1}\left( \frac{i}{\hbar}\left[  Q_{\hbar}(f_{1}),Q_{\hbar}(f_{2})\right]  \right)=\left\lbrace f_{1},f_{2}\right\rbrace \,,
\end{equation}
the latter known as Bohr's correspondence principle. The relation between the classical observable $f\in C^{\infty}(\mathbb{R}^{2})$ and its quantum counterpart $Q_{\hbar}(f_{})\in\mathfrak{A}$, in general does not correspond to an isomorphism of Lie algebras~\cite{Takhtajan}. 

In particular, the quantization mapping $Q_{\hbar}:\mathcal{A}\rightarrow \mathfrak{A}$, applied to a classical system described by the phase space $\mathbb{R}^{2}$, with local coordinates $p,q$ means the passage from the Poisson bracket which has the simple form
\begin{equation}
\left\lbrace q,p\right\rbrace=1 \,, 
\end{equation}
to the commutator of the operators
\begin{equation}\label{HCR}
\left[ \hat{Q},\hat{P}\right]\psi=i\hbar\psi\,, \;\;\; \textrm{where}\;\;\; \psi\in D\,, 
\end{equation}
where $D\subseteq\mathcal{H}$ is a dense subset of the Hilbert space $\mathcal{H}$, such that $\hat{Q}:D\rightarrow D$, and $\hat{P}:D\rightarrow D$. The equation (\ref{HCR}) is known as the Heisenberg commutation relation. The prescription $Q_{\hbar}(q)=\hat{Q}$ and $Q_{\hbar}(p)=\hat{P}$, satisfying (\ref{HCR}), provides the usual cornerstone for the quantization of most classical systems, and its validity results widely confirmed by numerous experiments.
In order to study the representation of the quantum kinematics in a particular Hilbert space, we consider the algebra generated by the operators given by the exponentiated versions of $\hat{Q}$ and $\hat{P}$, denoted by
\begin{equation}\label{Wgenerators}
\hat{U}(u)=e^{-iu\hat{P}/\hbar}, \;\;\;\; \hat{V}(v)=e^{-iv\hat{Q}/\hbar} \,,
\end{equation} 
where $u$ and $v$ are real parameters with dimensions of length and momentum, respectively. Whenever $\hbar\neq 0$, the operators $\hat{U}(u)$ and $\hat{V}(v)$ satisfy the commutation relation
\begin{equation}\label{Wrelation}
\hat{U}(u)\hat{V}(v)=e^{iuv/\hbar}\hat{V}(v)\hat{U}(u) \,.
\end{equation}
Then, the Weyl algebra denoted by $\mathcal{W}$, will be the algebra generated by finite linear combination of the operators $\hat{U}(u)$ and $\hat{V}(v)$, such that
\begin{equation}
\sum_{i}\left( a_{i}\hat{U}(u_{i})+b_{i}\hat{V}(v_{i})\right)\in\mathcal{W}\,, \;\;\;\textrm{where}\;\;\;a_{i},b_{i}\in\mathbb{C}\,. 
\end{equation}
From this point of view, defining a quantization mapping means to provide a unitary representation of the Weyl algebra on a Hilbert space.

In order to construct the Schr\"{o}dinger representation of the Weyl algebra in the Gaussian measure, let us select the Hilbert space to be
\beq
\mathcal{H}_{d}=L^{2}(\mathbb{R},d\mu_{d}) \,,
\eeq
given by the space of square integrable functions with respect to the Gaussian weighted measure on $\mathbb{R}$ 
\begin{equation}
d\mu_{d}(q)=\frac{1}{d\sqrt{\pi}}e^{-\frac{q^{2}}{d^{2}}}dq \,,
\end{equation}
where $d$ is a parameter with dimensions of length. In this Hilbert space, the position and momentum operators are represented as
\begin{equation}
\hat{Q}\psi(q)=(q\psi)(q) \;\;\;\textrm{and}\;\;\, \hat{P}\psi(q)=-i\hbar\frac{\partial}{\partial q}\psi(q)+i\hbar\frac{q}{d^{2}}\psi(q)\,. 
\end{equation}
This unusual expression for the representation of the momentum comes from the fact that in the Gaussian measure the momentum operator requires an extra term in order to define a symmetric operator or, in case we have identified a dense domain, the corresponding self-adjoint operator. 
The main reason to choose this particular representation relies on the Gelfand-Naimark-Segal (GNS) construction in which the 
algebraic structure of a complete set of quantum observables solely characterizes the Hilbert space. As it is well known, the GNS description is totally equivalent to providing a quantum representation in terms of bounded operators and probability measures~\cite{Strocchi}. 
The different representations of the Weyl algebra that can be obtained are trivialized by the Stone-von Neumann uniqueness theorem which asserts that all regular irreducible representations are unitarily equivalent. In this manner, it is possible to recover the standard Schr\"{o}dinger representation in the Hilbert space $\mathcal{H}_{Schr}=L^{2}(\mathbb{R},dq)$ from $\mathcal{H}_{d}$. For this, we need to give an isometric isomorphism $T:\mathcal{H}_{d}\rightarrow\mathcal{H}_{Schr}$, defined by
\begin{equation}
\psi(q):=T\varphi(q)=\frac{1}{d^{1/2}\pi^{1/4}}e^{-\frac{q^{2}}{2d^{2}}}\varphi(q) \,, 
\end{equation}
where $\psi\in\mathcal{H}_{Schr}$ and $\varphi\in\mathcal{H}_{d}$.  In this sense, all the $d$-representations in $\mathcal{H}_{d}$ are unitarily equivalent by the Stone-von Neumann theorem~\cite{CTZ}, \cite{Folland}. 

Next, in order to find the Weyl transform, we start by defining   $\hat{S}(u,v)\in\mathcal{L}(\mathcal{H}_{d})$ as a linear operator on $\mathcal{H}_{d}$ given by
\begin{equation}\label{Suv}
\hat{S}(u,v):=e^{\frac{-iuv}{2\hbar}}\hat{U}(u)\hat{V}(v) \,,
\end{equation} 
and note from relation~(\ref{Wrelation}) 
that this operator follows the identities
\begin{equation}\label{SS}
\hat{S}(u_{1},v_{1})\hat{S}(u_{2},v_{2})=e^{\frac{i}{2\hbar}(u_{1}v_{2}-u_{2}v_{1})}\hat{S}(u_{1}+u_{2},v_{1}+v_{2}) \,,
\end{equation}
and
\begin{equation}
\label{eq:SComplexConj}
\hat{S}(u,v)^{\dagger}=\hat{S}(-u,-v)  \,,
\end{equation}
where here the dagger symbol in the left-hand side 
of~(\ref{eq:SComplexConj}) stands for the adjoint of the operator $\hat{S}(u,v)$.   
Now, let us define a linear map $W:L^{1}(\mathbb{R}^{2})\rightarrow\mathcal{L}(\mathcal{H}_{d})$, called the Weyl transform, as
\begin{equation}\label{Wtransform}
W(f)=\frac{1}{2\pi\hbar}\int_{\mathbb{R}^{2}}f(u,v)\hat{S}(u,v)dudv \,,
\end{equation}
where the function $f(u,v)$ is defined on the classical phase space and the integral should be understood in a weak sense, that is,
for every $\phi_{1}$, $\phi_{2}\in\mathcal{H}_{d}$ the inner product
\begin{equation}
\left\langle W(f)\phi_{1},\phi_{2}\right\rangle_{\mathcal{H}_{d}}=\int_{\mathbb{R}^{2}}f(u,v)\left\langle \hat{S}(u,v)\phi_{1},\phi_{2}\right\rangle_{\mathcal{H}_{d}} \frac{dudv}{2\pi\hbar}
\end{equation}
is absolutely convergent and determines a bounded operator $W(f)$ which satisfies
\begin{equation}
||W(f)||\leq \frac{1}{2\pi\hbar}||f||_{L^{1}} \,.
\end{equation}
Using the properties of the Weyl algebra, one may show that the 
Weyl transform (\ref{Wtransform}) meets the properties
\begin{enumerate}
\item For all $f\in L^{1}(\mathbb{R}^{2})$,
\beq
\nn
W\left(f(u,v) \right)^{\dagger}=W\left(\overline{f(-u,-v)} \right) \,. 
\eeq
\item $\ker{W}=\{0\}$.
\item There is a homomorphism between $L^{1}(\mathbb{R})$ and $\mathcal{L}(\mathcal{H}_{d})$ given by
\begin{equation}\label{star}
W(f_{1})W(f_{2})=W(f_{1}\ast f_{2}) \;\;\;\textrm{for all}\;\;\; f_{1},f_{2}\in L^{1}(\mathbb{R}^{2}) \,,
\end{equation}
where the star-product is given by
\begin{equation}
\label{eq:StarP}
\mkern-60mu (f_{1}\ast f_{2})(u,v)=\frac{1}{2\pi\hbar}\int_{\mathbb{R}^{2}}e^{\frac{i}{2\hbar}(uv'-u'v)}f_{1}(u-u',v-v')f_{2}(u',v')du'dv \,.
\end{equation}
\end{enumerate}  
The first property follows from the definition of the Weyl transform (\ref{Wtransform}),  while the second property is a direct result from the completeness of the Hilbert space $\mathcal{H}_{d}$. In order to prove the third assertion, we invoke property (i) and relation (\ref{SS}), and observe 
that within the inner product we have
\begin{eqnarray}
\mkern-60mu \left\langle W(f_{1})W(f_{2})\phi_{1},\phi_{2}\right\rangle_{\mathcal{H}_{d}}&=& \left\langle W(f_{2})\phi_{1},W(f_{1})^{\dagger}\phi_{2}\right\rangle_{\mathcal{H}_{d}} \nonumber \\
&=&\frac{1}{2\pi\hbar}\int_{\mathbb{R}^{2}}f_{2}(u_{2},v_{2})\left\langle \hat{S}(u_{2},v_{2})\phi_{1},W(f_{1})^{\dagger}\phi_{2}\right\rangle_{\mathcal{H}_{d}}du_{2}dv_{2} \nonumber\\
&=& \frac{1}{2\pi\hbar}\int_{\mathbb{R}^{2}}\left(f_{1}\ast f_{2} \right)(u,v)\left\langle \hat{S}(u,v)\phi_{1},\phi_{2}\right\rangle_{\mathcal{H}_{d}}dudv \,,  
\end{eqnarray}
where $f_{1}\ast f_{2}\in L^{1}(\mathbb{R}^{2})$.  We also note that whenever we consider $\hbar=0$ we recover the ordinary convolution product between integrable functions~\cite{Takhtajan}.

\subsection{The Wigner-Weyl quantization}
\label{ssec:WWquantization}
In order to extend the Weyl transform to a more general class of functions, such as distributions, it is convenient to introduce the following linear map
\begin{equation}\label{Wquantizer}
\Phi:=W\circ\mathfrak{F}^{-1}:\mathcal{S}(\mathbb{R}^{2})\rightarrow\mathcal{L}(\mathcal{H}_{d}) \,.
\end{equation} 
According to this definition, $\Phi$ is a linear map from the Schwartz space $\mathcal{S}(\mathbb{R}^{2})$ of complex valued functions whose derivatives are rapidly decreasing into the linear operator space $\mathcal{L}(\mathcal{H}_{d})$.   This map
is   
called the Weyl quantization. Here $W$ corresponds to the Weyl transform defined in (\ref{Wtransform}) and $\mathfrak{F}^{-1}$ stands for the inverse Fourier transform
\begin{equation}
\tilde{f}(u,v)=\mathfrak{F}^{-1}(f)(u,v)=\frac{1}{2\pi\hbar}\int_{\mathbb{R}^{2}}f(p,q)e^{\frac{i}{\hbar}(up+vq)}dpdq  \,.
\end{equation}\label{Wmap}
By using the explicit expression for $\hat{S}(u,v)$ in (\ref{Suv}) and the inverse Fourier transformation, the Weyl quantization map reads
\begin{equation}\label{eq:WeylMap}
\Phi(f)\varphi(q)=\frac{1}{2\pi\hbar}\int_{\mathbb{R}^{2}}f\left(p,\frac{q+q'}{2}\right)e^{\frac{i}{\hbar}p(q-q')}e^{-\frac{1}{2d^{2}}(q^{2}-q'^{2})}\varphi(q')dpdq' \,, 
\end{equation}  
for $\varphi\in\mathcal{H}_{d}$. This means that the operator $\Phi$ represents an integral operator acting on $\mathcal{H}_{d}$
\begin{equation}
\Phi(f)\varphi(q)=\int_{\mathbb{R}}K(q,q')\varphi(q')dq' \,,
\end{equation} 
where the kernel $K(q,q')$ is given by
\begin{equation}\label{Kernel}
K(q,q')=\frac{1}{2\pi\hbar}\int_{\mathbb{R}}f\left(p,\frac{q+q'}{2}\right)e^{\frac{i}{\hbar}p(q-q')}e^{-\frac{1}{2d^{2}}(q^{2}-q'^{2})}dp \,.
\end{equation}
Since the Fourier transform maps the Schwartz space onto itself, this implies that the Weyl quantizer turns out to be a Hilbert-Schmidt operator acting on $\mathcal{H}_{d}$, that is, an operator with a well defined trace, but possibly infinite~\cite{ReedI}.
The inverse map $\Phi^{-1}=\mathfrak{F}\circ W^{-1}$ associated to the Weyl quantizer, also known as the Weyl's inversion formula, can be obtained as
\begin{equation}\label{Weylinversion}
f(p,q)=\mathfrak{F}\left(\Tr({\Phi(f)}S(u,v)^{-1}) \right)\,, 
\end{equation} 
where $f\in\mathcal{S}(\mathbb{R}^{2})$ and the trace is taken along an orthonormal basis for $\mathcal{H}_{d}$~\cite{Compean}. By using the kernel given in (\ref{Kernel}), the explicit expression for the Weyl's inversion formula reads
\begin{equation}\label{Winv}
f(p,q)=\int_{\mathbb{R}}K\left(q+\frac{z}{2},q-\frac{z}{2} \right)e^{-\frac{i}{\hbar}z(p-\frac{i\hbar q}{d^{2}})}dz \,. 
\end{equation}

Bearing this in mind, it is possible to define the Wigner function, which corresponds to a phase space representation of a quantum state in the following manner. Let $\hat{\rho}$ be a density operator associated to a quantum state $\varphi\in\mathcal{H}_{d}$, that is, a self-adjoint, positive semi-definite operator of trace one  written as 
\begin{equation}\label{density}
\hat{\rho}\phi(q)=\varphi(q)\int_{\mathbb{R}}\overline{\varphi(q')}\phi(q')d\mu_{d}(q') \,,
\end{equation} 
(or $\hat{\rho}=\ket{\varphi}\bra{\varphi}$ in Dirac notation), where $\phi, \varphi\in\mathcal{H}_{d}$. From (\ref{density}), we can observe that the operator $\hat{\rho}$ is an integral operator, then by the Weyl's inversion formula (\ref{Winv}) its corresponding phase space representation function is given by
 \begin{equation}\label{Wigner}
\rho(p,q)=\int_{\mathbb{R}}\varphi\left(q+\frac{z}{2}\right)\overline{\varphi\left(q-\frac{z}{2} \right)}e^{-\frac{i}{\hbar}zp} e^{-\frac{1}{d^{2}}(q^{2}+\frac{z^{2}}{4})}\frac{dz}{d\sqrt{\pi}}\,.  
 \end{equation}
This is the Wigner function and, as one may easily check, it is normalized $\frac{1}{2\pi\hbar}\int_{\mathbb{R}^{2}}\rho(p,q)dpdq=1$, 
whenever we consider normalized wave functions  
$\varphi \in\mathcal{H}_{d}$ on its definition~(\ref{Wigner}). Further, the projections on the momentum and position results in marginal probability densities 
\begin{eqnarray}
\frac{1}{2\pi\hbar}\int_{\mathbb{R}}\rho(p,q)dp
& = &
||\varphi||^{2}_{\mathcal{H}_{d}} \,, \nonumber\\
\frac{1}{2\pi\hbar}\int_{\mathbb{R}}\rho(p,q)dq
& = & 
||\mathfrak{F}\left(T\varphi\right)||^{2}_{\mathcal{H}_{Schr}} \,. 
\end{eqnarray}
Another important property of the Wigner function lies on the possibility to take negative values in  
certain regions of  phase 
space. This last quasi-distributional aspect provides a quantum  device to measure interference and entanglement using only classical and statistical features~\cite{Curtright}.
To finish this section, we should recall that given an arbitrary operator $\hat{A}\in\mathcal{L}(\mathcal{H}_{d})$, the Wigner function can be used to obtain the expectation value 
of that operator, $\braket{\hat{A}}_{\mathcal{H}_{d}}$,  as a phase space average (in a particular ordering prescription) by
\begin{equation}\label{expectation}
\frac{1}{2\pi\hbar}\int_{\mathbb{R}^{2}}\rho(p,q)A(p,q)dpdq=\left\langle\varphi,\hat{A}\varphi\right\rangle_{\mathcal{H}_{d}} \,,  
\end{equation}     
where the operator $\hat{A}=\Phi(A)$ corresponds to the Weyl transform of the classical phase space function $A(p,q)$.

\section{The Polymer Representation}
\label{sec:Polymer}
\subsection{Polymer Quantum Mechanics}
In this section we derive the polymer representation of quantum mechanics from the standard Schr\"{o}dinger representation within the formalism of Wigner-Weyl quantization developed in the previous section. We will first analyze the Weyl algebra by means of the Weyl transform, and then through some distributional limits we will obtain the Wigner function and the 
star-product associated to the polymer representation. 
As a first step, we require to establish how the Weyl algebra $\mathcal{W}$, generated by $\hat{U}(u)$ and $\hat{V}(v)$ is represented on $\mathcal{H}_{d}=L^{2}(\mathbb{R},d\mu_{d})$. By using the Weyl map (\ref{eq:WeylMap}), we obtain
\begin{eqnarray}\label{WeylU}
\mkern-60mu\Phi(U(u))\varphi(q)&=&\frac{1}{2\pi\hbar}\int_{\mathbb{R}^{2}}U\left(p,\frac{q+q'}{2}\right)e^{\frac{i}{\hbar}p(q-q')}e^{-\frac{1}{2d^{2}}(q^{2}-q'^{2})}\varphi(q')dpdq'\,,\nonumber\\
&=&e^{\frac{u}{d^{2}}(q-\frac{u}{2})}\varphi(q-u)\,, 
\end{eqnarray}
and
\begin{eqnarray}\label{WeylV}
\mkern-60mu\Phi(V(v))\varphi(q)&=&\frac{1}{2\pi\hbar}\int_{\mathbb{R}^{2}}V\left(p,\frac{q+q'}{2}\right)e^{\frac{i}{\hbar}p(q-q')}e^{-\frac{1}{2d^{2}}(q^{2}-q'^{2})}\varphi(q')dpdq'\,,\nonumber\\
&=&e^{-\frac{i}{\hbar}vq}\varphi(q)\,. 
\end{eqnarray}

In this representation, obtained via the GNS construction, the vacuum state is given by the identity function $\varphi_{0}(q)=1$. The vacuum expectation value corresponding to the generators of the Weyl algebra, $\hat{U}(u)$ and $\hat{V}(v)$, may be realized by using (\ref{expectation}) with $\hat{\rho}_{0}=\ket{\varphi_{0}}\bra{\varphi_{0}}$, and explicitly reads
\begin{equation}
\left\langle\varphi_{0},\hat{U}(u)\varphi_{0} \right\rangle_{\mathcal{H}_{d}}=\frac{1}{2\pi\hbar}\int_{\mathbb{R}^{2}}\rho_{0}(p,q)U(u)dpdq=e^{-\frac{u^{2}}{4d^{2}}}  \,,
\end{equation} 
and
\begin{equation}
\left\langle\varphi_{0},\hat{V}(v)\varphi_{0} \right\rangle_{\mathcal{H}_{d}}=\frac{1}{2\pi\hbar}\int_{\mathbb{R}^{2}}\rho_{0}(p,q)V(v)dpdq=e^{-\frac{v^{2}d^{2}}{4\hbar^{2}}} \,.
\end{equation}
where $\rho_0(p,q)$ may be obtained by explicitly inserting the
states $\varphi_0(q)=1$ in the Wigner function~(\ref{Wigner}), thus yielding
\beq
\rho_0(p,q)=\int_{\mathbb{R}}e^{-\frac{i}{\hbar}zp} e^{-\frac{1}{d^{2}}(q^{2}+\frac{z^{2}}{4})}\frac{dz}{d\sqrt{\pi}}\,.  
\eeq  
We can observe that the representation of the Weyl algebra in the Hilbert space $\mathcal{H}_{d}$, for $d>0$, is well defined and continuous in any value of $u$ and $v$.

Following \cite{CTZ}, our purpose now is to obtain the polymer representation as a distributional limit from the Weyl algebra evaluated on $\mathcal{H}_{d}$. The main idea is to study two possible limits for the parameter $d$, the limit $d\rightarrow 0$, and the limit $1/d\rightarrow 0$. Contrary to the cases analyzed in \cite{CTZ}, where the states become ill-defined and the heuristic introduction of square roots of Delta distributions as half densities is required, within the Wigner-Weyl formalism both limits are well defined and the resulting representation will allow us to obtain the corresponding Wigner function associated with Loop Quantum Cosmology.
Before proceeding, it is convenient to focus on the fundamental vector states belonging to the Hilbert space $\mathcal{H}_{d}$, that is, the vectors generated by the action of the Weyl algebra generators $\hat{U}(u)$ and $\hat{V(v)}$ on the vacuum state $\varphi_{0}$. Let us call them
\begin{equation}\label{phiu}
\phi_{u}(q):=\hat{U}(u)\varphi_{0}(q)=\Phi(U(u))\varphi_{0}(q)=e^{\frac{u}{d^{2}}(q-\frac{u}{2})} \,,
\end{equation}
and
\begin{equation}\label{varphiv}
\varphi_{v}(q):=\hat{V}(v)\varphi_{0}(q)=\Phi(V(v))\varphi_{0}(q)=e^{-\frac{i}{\hbar}vq}\,,
\end{equation}
respectively, where we have used the fact that $\varphi_0(q)=1$ in each of the last identities.
Now denoting by $\hat{\rho}_{\phi tu}:=\ket{
\phi_{u}}\bra{\phi_{t}}$ and $\hat{\rho}_{\varphi vw}:=\ket{
\varphi_{v}}\bra{\varphi_{w}}$ and inserting these in expression (\ref{expectation}), we find that the inner product between these states is given by
\begin{equation}\label{innerU}
\left\langle \phi_{u},\phi_{s}\right\rangle_{\mathcal{H}_{d}}=\frac{1}{2\pi\hbar}\int_{\mathbb{R}^{2}}\rho_{\phi us}dpdq=e^{-\frac{1}{4d^{2}}(u-s)^{2}} \,, 
\end{equation}
and
\begin{equation}\label{innerV}
\left\langle \varphi_{v},\varphi_{w}\right\rangle_{\mathcal{H}_{d}}=\frac{1}{2\pi\hbar}\int_{\mathbb{R}^{2}}\rho_{\varphi vw}dpdq=e^{-\frac{d^{2}}{4\hbar^{2}}(v-w)^{2}} \,, 
\end{equation}
where the Wigner functions $\rho_{\phi us}$ and $\rho_{\varphi vw}$ are obtained by introducing the states~(\ref{phiu}) and~(\ref{varphiv}) in~(\ref{Wigner}), respectively. 
Thus we can observe that, unlike the Schr\"{o}dinger representation, in the Hilbert space $\mathcal{H}_{d}$ plane waves are normalized.
With the preceding calculations, we start by analyzing the limit $1/d\mapsto 0$. From the inner product (\ref{innerU}) and (\ref{innerV}) we get
\begin{equation}
\lim_{1/d\mapsto 0}\left\langle \phi_{u},\phi_{s}\right\rangle_{\mathcal{H}_{d}}= 1 \,,
\end{equation}
and
\begin{equation}\label{innervarphi}
\lim_{1/d\mapsto 0}\left\langle \varphi_{v},\varphi_{w}\right\rangle_{\mathcal{H}_{d}}=\delta_{v,w} \,,
\end{equation} 
where $\delta_{v,w}$ stands for a Kronecker delta, implying that the states $\varphi_{v}(q)$ form an orthonormal basis for the new Hilbert space obtained as a limiting case from $\mathcal{H}_{d}$ quotient the vector space  generated by $\phi_{u}(q)$ since any difference of these vectors has zero norm. It is important to note that within this limit, on the one hand the operators $\hat{V}(v)$ and the momentum operator $\hat{p}$ are well defined, as one may straightforwardly check by using the Weyl map
\begin{eqnarray}
\mkern-60mu \hat{U}(s)\varphi_{v}(q)=\Phi(U(s))\varphi_{v}(q)&=&\frac{1}{2\pi\hbar}\int_{\mathbb{R}^{2}}e^{-\frac{i}{\hbar}sp}e^{\frac{i}{\hbar}p(q-q')}e^{-\frac{1}{2d^{2}}(q^{2}-q'^{2})}e^{-\frac{i}{\hbar}vq'}dpdq' \nonumber\\
&=& e^{\frac{s}{d^{2}}(q-\frac{s}{2})}e^{-\frac{i}{\hbar}v(q-s)} \,, 
\end{eqnarray} 
which in the limit $1/d\mapsto 0$ reduces to
\begin{equation}
\hat{U}(s)\varphi_{v}(q)\mapsto e^{\frac{i}{\hbar}sv}\varphi_{v}(q) \,,
\end{equation}
while for the momentum operator we have 
\begin{eqnarray}\label{pgauss}
\hat{P}\varphi_{v}(q)=\Phi(p)\varphi_{v}(q)&=&\frac{1}{2\pi\hbar}\int_{\mathbb{R}^{2}}pe^{\frac{i}{\hbar}p(q-q')}e^{-\frac{1}{2d^{2}}(q^{2}-q'^{2})}e^{-\frac{i}{\hbar}vq'}dpdq'     \nonumber\\
&=& -i\hbar\frac{d\varphi_{v}}{dq}+\frac{i}{\hbar}\frac{q}{d^{2}}\varphi_{v} \,, 
\end{eqnarray} 
which in the limit $1/d\mapsto 0$ behaves as
\begin{equation}\label{Amomentum}
\hat{P}\varphi_{v}(q)\mapsto -i\hbar\frac{d\varphi_{v}}{dq}  \,.
\end{equation}
On the other hand, the operator $\hat{V}(w)$ acts on the fundamental vector states as
\begin{equation}
\hat{V}(w)\varphi_{v}(q)=\Phi(V(w))\varphi_{v}(q)=e^{-\frac{i}{\hbar}wq}\varphi_{v}(q) \,,
\end{equation}
which can be easily computed by formula (\ref{WeylV}). Nevertheless, we may notice that within this representation there is no  position operator $\hat{Q}$ as a consequence of the fact 
that the operator $\hat{V}(w)$
is not weakly continuous, that is,  the inner product 
\begin{equation}
\left\langle\varphi_{v},\hat{V}(w)\varphi_{y}\right\rangle\mapsto\delta_{w,v-y} \,, 
\end{equation} 
is not continuous on the parameter $w$.  Indeed, if one considers the  
basis $\{\varphi_v\}$ one may straightforwardly check 
that within this inner product the
expectation value  of the position operator $\hat{Q}$ 
always vanishes in the $1/d\mapsto 0$ limit.
This means, by the Stone-von Neumann theorem, that the present limit representation is not equivalent to the standard Schr\"{o}dinger representation since we have broken the regularity condition. The resulting Hilbert space is known in the literature as the A-version of the polymer representation~\cite{CTZ}. 

Let us now analyze the Hilbert space $\mathcal{H}_{d}$ in the limit $d\mapsto 0$.  Contrary to the previous case we obtain from the inner products (\ref{innerU}) and (\ref{innerV}) 
\begin{equation}\label{innerphi}
\lim_{d\mapsto 0}\left\langle \phi_{u},\phi_{s}\right\rangle_{\mathcal{H}_{d}}=\delta_{u,s} \,,
\end{equation}
and
\begin{equation}
\lim_{d\mapsto 0}\left\langle \varphi_{v},\varphi_{w}\right\rangle_{\mathcal{H}_{d}}= 1\,,
\end{equation} 
respectively. 
We recognize that within this limit the vectors $\phi_{u}$ become the orthonormal basis and, in order to obtain a proper Hilbert space, we have to take the quotient of the vector space generated by the functions $\varphi_{v}$. 
From (\ref{pgauss}), we readily see that in this representation the momentum operator is not going to be defined. In order to formally establish this result we only require to calculate the inner product
\begin{equation}
\left\langle \phi_{s},\hat{U}(u)\phi_{t}\right\rangle\mapsto\delta_{u,s-t} \,,
\end{equation}
which occurs to be not weakly continuous on the parameter $u$, and as a consequence of the Stone-von Neumann theorem, the resulting $d\mapsto 0$ limit of the Hilbert space $\mathcal{H}_{d}$ brings another inequivalent representation, called the B-version of the polymer representation~\cite{CTZ}. 

For both limiting Hilbert spaces, the A-polymer and the B-polymer versions, although very similar, it is important to emphasize the
distinct roles played by the generators of the Weyl algebra, $\hat{U}(u)$ and $\hat{V}(v)$, in each of these cases. In the A-polymer representation, the position operator $\hat{Q}$ is not well defined since $\hat{V}(v)$ is not weakly continuous, and conversely, in the case of the B-polymer representation the momentum operator $\hat{P}$ is not well defined due to the lack of weak continuity on the operator $\hat{U}(u)$.
Having computed both limits, one can find an equivalence relation between the A and B polymer representation through a $d \leftrightarrow 1/d$ duality between the position and momentum representations.  The A-polymer case in the position representation turns out to be equivalent to the B-polymer version in the momentum representation and vice versa. In order to see this, it is necessary to analyze the fundamental states of $\mathcal{H}_{d}$ in both limits. Nevertheless, as we can observe from expression (\ref{phiu}), the states $\phi_{u}$ become ill-defined in the $d\mapsto 0$ limit.   To overcome these difficulties, in reference~\cite{CTZ} the authors considered to incorporate these states into the standard Schr\"{o}dinger representation, but this choice has the consequence of introducing square roots the of Dirac delta distributions, which are not well defined since Schwartz-Sobolev distributions are linear functionals on the space of test functions \cite{Kanwal}. In order to avoid the introduction of non-linear distributions by means of heuristic arguments, in the next section we will investigate the properties of the Wigner function and its polymer limits.  

\subsection{The polymer Wigner function and the star-product}              

Let us now analyze the Wigner function corresponding to the A and B-polymer representations as limiting cases of the Wigner function associated to the Hilbert space $\mathcal{H}_{d}$. The resulting Wigner functions associated with the limits $1/d\mapsto 0$ and $d\mapsto 0$ will be called $\rho_{A}(p,q)$ and $\rho_{B}(p,q)$, respectively. Starting with the Wigner function defined in (\ref{Wigner}) for the vector states $\varphi_{v}=e^{-\frac{i}{\hbar}vq}$ 
\begin{equation}\label{WignerPoly1}
\rho_{\varphi_v}(p,q)=\int_{\mathbb{R}}\varphi_{v}\left(q+\frac{z}{2}\right)\overline{\varphi_{v}\left(q-\frac{z}{2} \right)}e^{-\frac{i}{\hbar}zp} e^{-\frac{1}{d^{2}}(q^{2}+\frac{z^{2}}{4})}\frac{dz}{d\sqrt{\pi}} \,, 
 \end{equation}
and then taking the limit $1/d\mapsto 0$ for the Wigner function $\rho_{\varphi}(p,q)$, we obtain the explicit expression for the Wigner function corresponding to the A-polymer representation
\begin{equation}\label{rhoA}
\rho_{\varphi_v}(p,q)\mapsto \delta_{p,-v}=:\rho_{A}(p,q)  \,.
\end{equation} 
Similarly, for the limiting case $d\mapsto 0$ associated to the Wigner function defined by the vector states $\phi_{u}=e^{\frac{u}{d^{2}}(q-\frac{u}{2})}$
\begin{equation}\label{WignerPoly2}
\rho_{\phi_u}(p,q)=\int_{\mathbb{R}}\phi_{u}\left(q+\frac{z}{2}\right)\overline{\phi_{u}\left(q-\frac{z}{2} \right)}e^{-\frac{i}{\hbar}zp} e^{-\frac{1}{d^{2}}(q^{2}+\frac{z^{2}}{4})}\frac{dz}{d\sqrt{\pi}} \,, 
 \end{equation}
we obtain the Wigner function corresponding to the B-polymer representation
\begin{equation}\label{rhoB}
\rho_{\phi_u}(p,q)\mapsto \delta_{q,u}=:\rho_{B}(p,q) \,.
\end{equation}
These ultra-localized expressions for the quasi-probability distributions in the phase space given by $\rho_{A}$ and $\rho_{B}$, are distinctive in the polymer representation of  quantum mechanics since the wave functions are modulated by Kronecker deltas on a countable (or possibly uncountable) number of points.  However, we may note 
that the true nature of the lattice formed by these points is not 
explicitly obtained, and thus, for simplicity,  we will consider a regular 
lattice from now on. This means that the polymer Hilbert space consists of wave functions that vanish everywhere except for a countable number of points on the real line which are regularly spaced $q_{n}=q_{0}+n\lambda$, for a given $q_{0}\in\mathbb{R}$ and $n\in\mathbb{Z}$. For a fixed point $q_0$,  the wave functions, supported on this lattice, belong to a separable Hilbert space which corresponds to a superselected sector of the full polymer Hilbert space. Indeed, a state defined on this space is written as a linear superposition of all the functions defined on the lattices indexed by $q_{0}$, where $q_{0}\in [0,\lambda)$. Hence, the polymer Hilbert space  is given by a direct sum of the superselected sectors $\mathcal{H}_{q_{0}}$, i.e., $\mathcal{H}_{poly}=\oplus_{q_{0}\in [0,\lambda)}\mathcal{H}_{q_{0}}$.  This emergence of these superselection 
sectors is 
completely analogous to the situation appearing in the 
canonical construction of the polymer representation~\cite{Propagators}. Furthermore, the Wigner functions $\rho_{A}$ and $\rho_{B}$ calculated in (\ref{rhoA}) and (\ref{rhoB}), respectively, correspond to the Wigner functions associated to the pure characters and their Fourier transforms over the Bohr compactification $\mathbb{R}_{B}$ of the real line  in Loop Quantum Cosmology \cite{Sahlmann}. This means that both limits $1/d\mapsto 0$ and $d\mapsto 0$ of the Wigner function in $\mathcal{H}_{d}$ converge to
\begin{eqnarray}\label{polyA}
\rho_{A}(p,q) &=& \lim_{1/d\to 0}\rho_{\varphi_v}(p,q)=\int_{\mathbb{R}_{B}}\varphi_{v}\left( q+\frac{1}{2}b\right) \overline{\varphi_{v}\left( q-\frac{1}{2}b\right)}{h(b,-p)}db \,,\\
\label{polyB}
\rho_{B}(p,q)& = &\lim_{d\to 0}\rho_{\phi_v}(p,q)=\int_{\hat{\mathbb{R}}_{B}}\tilde{\varphi}_{v}\left( p-\frac{1}{2}\tau\right) \overline{\tilde{\varphi}_{v}\left( p+\frac{1}{2}\tau\right)}h(q,\tau)d\tau,
\end{eqnarray}
where $\mathbb{R}_{B}$ stands for the Bohr compactification over the reals, while 
$\hat{\mathbb{R}}_{B}$ stands for its locally compact dual group.  Also, $h(b,p)=e^{ip b}$ are the characters of $\mathbb{R}_{B}$ and $\tilde{\varphi}_{v}$ denotes the Fourier tranform of $\varphi_{v}$ in $\mathbb{R}_{B}$. In the context of LQC one can fix the gauge and diffeomorphism freedom of the full gravitational theory in such a way that the resulting phase space occurs to be finite dimensional. This means that the canonical variables given by the connection and the triad are parametrized by $(c,p_{c})$, where $c$ represents the configuration variable corresponding to the connection associated to fiducial edges, and $p_{c}$ represents its canonically conjugate momentum. In terms of geometrodynamical variables, $p_{c}$ provides the scale factor (which establishes the spatial metric in the case of homogeneous and flat cosmologies) and $c$ determines the extrinsic curvature \cite{Robustness}. Since a general wave function in the kinematical Hilbert space of LQC is given as a finite span of the fundamental characters in $\mathbb{R}_{B}$ (in other terms, a wave function  corresponds to a cylindrical function), the limits $1/d \to 0$ and $d\to 0$ correspond to the Wigner function within the context of Loop Quantum Cosmology in the position and momentum representations, respectively, 
characterizing the $1/d\leftrightarrow d$ duality between the A and B-polymer representations mentioned above. As we can observe from (\ref{polyA}) and (\ref{polyB}), the A-polymer representation in the position space is equivalent to the B-polymer representation in the momentum space, and vice versa. On the other hand, since the fundamental vector states $\phi_{u}$ and $\varphi_{v}$ are constructed by applying the operators $\hat{U}(u)$ and $\hat{V}(v)$, respectively, over the vacuum state $\varphi_{0}$, which by the GNS construction corresponds to a cyclic vector. This means that any other vector of the Hilbert space $\mathcal{H}_{d}$ can be obtained by applying a finite linear combination of the operators $\hat{U}(v)$ and $\hat{V}(v)$ on $\varphi_{0}$, as the representation of the Weyl algebra is dense on $\mathcal{H}_{d}$. Then, in order that the Wigner-Weyl map and the Wigner function result well defined under the $d\mapsto 0$ and $1/d \mapsto 0$ limits, the domain of $\Phi$ corresponds to the almost periodic functions on the phase space, that is, the space of finite linear combinations of sine and cosine trigonometric functions closed under the uniform norm~\cite{Rudin}. As a consequence of the fact that the integral expression of the Wigner function (\ref{rhoB})  with 
respect to the Hilbert space $\mathcal{H}_{d}$ contains also a Gaussian measure factor $d\mu_{d}$, it is possible to demonstrate that the limit of this kind of integrals converge to the normalized Haar measure on the Bohr compactification $\mathbb{R}_{B}$ with respect to the weak topology~\cite{Hormann},~\cite{Blum}. This analysis suggests that, within the deformation quantization approach introduced here, the polymer representation and therefore the Wigner function associated to LQC can be thought as a distributional limit of the standard Schr\"{o}dinger representation.

Let us now introduce the star-product associated with the polymer representation. The Wigner-Weyl quantization mapping $\Phi:\mathcal{S}(\mathbb{R}^{2})\rightarrow \mathcal{L}(\mathcal{H}_{d})$ studied in the previous section, defines a bilinear operation: $\ast:\mathcal{S}(\mathbb{R}^{2})\times \mathcal{S}(\mathbb{R}^{2})\rightarrow\mathcal{S}(\mathbb{R}^{2})$ given by $f_{1}\ast f_{2}=\Phi^{-1}\left(\Phi(f_{1})\Phi(f_{2}) \right)$ called the star-product, and written explicitly as (\ref{eq:StarP}). Even though this star-product is originated through a homomorphism between the spaces $\mathcal{S}(\mathbb{R}^{2})$ and $\mathcal{L}(\mathcal{H}_{d})$, one may easily verify that it does not depend on the parameter $d$.  This means that after performing the limits $1/d\mapsto 0$ or $d\mapsto 0$ the star-product will remain unchanged. In order to illustrate this, we briefly outline the differential representation of the star-product. Let $\hat{f}_{1}, \hat{f}_{2}\in\mathcal{L}(\mathcal{H}_{d})$, such that $\hat{f}_{1}=\Phi(f_{1})$ and $\hat{f}_{2}=\Phi(f_{2})$.  By Weyl's inversion formula (\ref{Weylinversion})

\begin{equation}
\Phi^{-1}(\hat{f}_{1}\hat{f}_{2})=\int_{\mathbb{R}^{2}}K_{f_{1}}\left(q+\frac{z}{2},y\right)K_{f_{2}}\left(y,q-\frac{z}{2}\right)e^{-\frac{i}{\hbar}z(p-\frac{i\hbar q}{d^{2}})}dzdy \,,
\end{equation}  
     
where $K_{f_{1}}$ and $K_{f_{2}}$ correspond to the kernel~(\ref{Kernel}) associated to the operators $\hat{f}_{1}$ and $\hat{f}_{2}$, respectively. To proceed further, we introduce a new set of variables $q'=y-q-\frac{z}{2}$, and $y'=q-y-\frac{z}{2}$, and consider the Taylor expansion of $K_{f_{1}}$ and $K_{f_{2}}$ and then, after some laborious manipulations~\cite{Teaching}  where we have to consider the Gaussian measure, we obtain
\begin{equation}\label{starproduct}
\left( f_{1}\ast f_{2}\right) (p,q)=f_{1}(p,q)\exp\left[-\frac{i\hbar}{2}(\overleftarrow{\partial_{q}}\overrightarrow{\partial_{p}}-\overleftarrow{\partial_{p}}\overrightarrow{\partial_{q}}) \right]f_{2}(p,q) \,. 
\end{equation}
The form of the star-product means that in the limiting cases $1/d\mapsto 0$ and $d\mapsto 0$, the correspondence principle
\begin{eqnarray}
\lim_{\hbar\to 0}\Phi^{-1}\left(\frac{i}{\hbar}\left[\hat{f}_{1},\hat{f}_{2} \right]\right)
& = &\lim_{\hbar\to 0}\frac{i}{\hbar}\left(f_{1}\ast f_{2}-f_{2}\ast f_{1} \right) \nonumber\\
& = &\left\lbrace f_{1},f_{2} \right\rbrace \,,   
\end{eqnarray} 	
is satisfied. Nevertheless, we must remember that not all operators are well defined in the polymer representation, for instance, the position operator is not defined in the A-polymer representation, while the momentum operator is not defined in the B-polymer representation. This shows that as long as we take into account phase space functions with well defined limits in each of the two polymer versions under the Wigner-Weyl map, the correspondence principle will remain satisfied.   

\section{The Uncertainty Principle }
In this section, we determine the uncertainty principle within the polymer representation. For simplicity, we only consider the case for the B-polymer prescription, though an analogous uncertainty principle will follow for the A-polymer representation. Let $f\in\mathcal{S}(\mathbb{R}^{2})$, then, as we learned in subsection~(\ref{ssec:WWquantization}), the expectation value of the operator $\hat{f}=\Phi(f)$ can be expressed as (\ref{expectation}). Since the star-product between classical functions is given by (\ref{starproduct}), and does not depend on the parameter $d$, it is easy to prove that $\braket{\overline{f}\ast f}\geq 0$ \cite{Curtright}. Also, following~\cite{Curtright}, in order to obtain the Heisenberg's uncertainty relation we need to choose 
\begin{equation}
f=a+bq+c\left(-\frac{1}{2iu}\left(U(u)-U(-u)\right)\right)\,, 
\end{equation}
where $a,b,c\in\mathbb{C}$ and the regularized momentum is defined in terms of the Weyl generators
\begin{equation}
p_u := -\frac{1}{2iu}\left(U(u)-U(-u)\right)=\frac{1}{u}\sin{up} \,,
\end{equation}
taking for simplicity $\hbar=1$. The reason for this definition  is a consequence of the B-polymer representation, since within this representation in the limit $d\mapsto 0$ the momentum operator does not exist as discussed above, and then the momentum has to be represented through the generators of the Weyl algebra. Although this election for the momentum is not unique, it is guided by the requirement of non-degenerate energy levels in a Hamiltonian with a quadratic kinetic term and, also, it has been proved useful in the study of the semiclassical regime which yields an effective dynamics related by coarse-graining maps \cite{physicalCorichi}. The expectation value results in a positive quadratic form 
\begin{eqnarray}
\mkern-60mu\braket{\overline{f}\ast f}&=&\overline{a}a+\overline{b}b\braket{q\ast q}+\overline{c}c\braket{p_u\ast p_u} +(\overline{a}b+\overline{b}a)\braket{q} +(\overline{a}c+\overline{c}a)\braket{p_u} 
\nonumber\\
& & +\overline{c}b\braket{p_u \ast q}+\overline{b}c\braket{q\ast p_u}\geq 0 \,.
\label{eq:expvalue}
\end{eqnarray}
Taking into account the usual definition of the quantum fluctuation of an operator $(\Delta f)^{2}=\braket{(f-\braket{f})^{2}}$, and the positivity condition of the quadratic form~(\ref{eq:expvalue}) (which results to be equivalent to the positivity of its associate $3\times 3$ matrix determinant) we can thus write
\beq
(\Delta q)^{2}(\Delta p_u)^{2} & \geq & 
\frac{1}{4}\langle \cos up \rangle^2
+\left\langle\left(q-\braket{q} \right)\left(p_u-\braket{p_u} \right)\right\rangle^{2} \nn\\
&\geq & 
\frac{1}{4}\langle \cos up \rangle^2  \,,
\label{eq:squareur}
\eeq  
and hence
\begin{equation}\label{polyur}
\Delta q\Delta p_u\geq 
\frac{1}{2} |\langle \cos up \rangle |
= \frac{1}{2}\langle \sqrt{1-u^2p_u^2} \rangle  \,.
\end{equation}
From this last expression it is easy to see that 
whenever we consider a small parameter $u$, we 
obtain the relation
\beq
\Delta q\Delta p_u\geq \frac{1}{2} \left( 1-\frac{1}{2}u^2\Delta p_u^2 + \mathcal{O}(u^4) \right)  \,,
\eeq
where we have considered the condition $\langle p_u \rangle=0$. This expression results exactly the uncertainty relation obtained
by~\cite{Husain}, which is completely 
analogous to the ones appearing within the context of generalized uncertainty principle (GUP) scenarios analyzed within String Theory and Loop Quantum Gravity~\cite{Kempf},~\cite{Nozari}, \cite{kho}. Nevertheless, in the case of the polymer representation there is a key difference with respect to the GUP's. Indeed, most of generalized uncertainty principle theories determine a nonvanishing minimal uncertainty in the position, $\Delta q_{0}>0$, such that $\Delta q\geq \Delta q_{0}$ (or analoguously, in the momentum), which implies, as discussed in~\cite{Kempf}, that under such minimal uncertainties, in the position (or the momentum) 
of a particle, there cannot be any physical state which is a position  (or momentum) eigenstate. 
However, as we can observe from expression (\ref{polyur}), the polymer approach does not imply any minimal uncertainty in the position. The discreteness properties within the polymer representation are encoded in the eigenvalues of the position and momentum operators. In order to observe this, let us take for example the polymer A-version. In this case, the vectors $\varphi_{v}(q)$ defined in (\ref{varphiv}) not only form an orthonormal basis, but they also correspond to eigenstates of the momentum operator $\hat{P}$ given in (\ref{Amomentum}), as
$
\hat{P}\varphi_{v}(q)=-v\varphi_{v}.
$
Since the eigenvalues of the momentum operator are $v\in\mathbb{R}$, this could suggest the appearance of a continuous spectrum. However, these eigenstates are also normalizable,  which comprise one of the characteristics of states belonging to the discrete spectrum. This contradiction is solved by taking into account the non-separability of the polymer Hilbert space $\mathcal{H}_{poly}$, where the normalizability condition on the eigenstates implies the discreteness of the spectrum.
It is important to mention that if one 
restricts to one superselection sector, then this sector turns out to be separable in opposition to the complete Hilbert space, and it also preserves a discrete spectrum.
Finally, in the context of LQC the kinematical Hilbert space is constructed by implementing the same steps   followed in the polymer representation. As we already mentioned in section III, in this case the phase space variables $(q,p)$ are given by the canonical pair $(c,p_{c})$, satisfying $\{c,p_{c}\}=8\pi\gamma G/3$, where $\gamma$ denote the Immirzi parameter and $G$ the Newton gravitational constant, respectively. Since the operator associated to the holonomy $e^{-i\mu c}$ is not weakly continuous, this means that the connection operator $\hat{c}$ is not defined. But this is exactly the A-version of the polymer representation, hence, this implies that the momentum operator $\hat{p}_{c}$, which is given by the densitized triad multiplied by a specific power of a volume of the region used to define the isotropic phase space, admits a discrete spectrum given by 
$
\hat{p}_{c}\varphi_{\mu}(c)=(8\pi\gamma/3)l_{p}^{2}\mu\varphi_{\mu}(c),
$
where $l_{p}$ is the Planck lenght, clearly this shows the discreteness properties of the triad spectrum. \cite{BojowaldLQC}
Similarly to the polymer case the curvature components, which are given by higher powers of the connection $c$, must be included in the quantization. This means that a kind of regularization must be selected, but in any case the resulting Hamiltonian constraint results in a difference equation rather than a differential equation. The action of this Hamiltonian operator superselects wave functions on $\mathcal{H}_{poly}$ which are preserved under dynamics. Although, it has been argued that the choice of superselection sector in LQC may be constructed by geometrically homogeneous quantum embeddings which could be identified with the symmetric sector of diffeomorphism invariant full Loop Quantum Gravity \cite{Koslowski},~\cite{Marugan}. We expect that the formalism developed here could shed some light on these issues by exploiting the technical tools that naturally emerge from the deformation quantization approach.

\section{Conclusions}
\label{sec:conclu}

In this paper, we showed that the non-regular polymer representation of quantum mechanics can be obtained as a distributional limit of the Schr\"{o}dinger representation within the deformation quantization approach.
The two limiting cases analyzed correspond to the Wigner function within the context of Loop Quantum Cosmology in the position and momentum representations, respectively.  Then, by using the polymer star-product obtained by the Weyl's inversion formula and taking the limit $\hbar\mapsto 0$, we recovered the standard Poisson structure for smooth functions, thus fulfilling Bohr's correspondence principle. Finally, we derived the uncertainty relations between the position and momentum operator. However, as we have seen, under the polymer representation that we have considered the momentum operator is not well defined,  meaning that an approximation in terms of Weyl generators must be taken. By considering the appropriate limits, our construction results analogous to the uncertainty relation that appears in the context of different generalized uncertainty principle scenarios. Nevertheless, in the case of the polymer representation the discreteness properties are encoded in the eigenvalues of the position and momentum operators.

The above mentioned properties suggest that the non-regular polymer representation described as an appropriate distributional limit in terms of Wigner quasi-probability functions not only follows the correspondence limit among quantum algebraic structures such as the commutator of self-adjoint operators and the classical Poisson bracket between smooth functions, but it also provides at the same time a minimal length scale at the quantum level. We expect the results established here may clarify the quantum dynamics of systems defined on independent background scenarios. However, a more general analysis must be carried out in order to apply the methods developed here for the case of field theories. This will be done elsewhere.

\section*{Acknowledgments}
The authors would like to acknowledge financial support from CONACYT-Mexico
under projects CB-2014-243433 and CB-2017-283838.

\section*{References}

\bibliographystyle{unsrt}

\begin{thebibliography}{l}

\bibitem{shadow} A.~Ashtekar, S.~Fairhurst and J.~L.~Willis, \emph{Quantum gravity, shadow states, and quantum mechanics}, Class.~Quantum~Grav. {\bf 20} 1031--1062 (2003), \texttt{arXiv:gr-qc/0207106}.

\bibitem{CTZ}A.~Corichi, T.~Vukasinac and J.~A.~Zapata, \emph{Polymer Quantum Mechanics and its Continuum Limit}, Phys.~Rev. {\bf D76}, 044016 (2007), \texttt{arXiv:0704.0007v2 [gr-qc]}.

\bibitem{Bojowald} M.~Bojowald, \emph{Loop quantum cosmology}, Living Rev.~Relativ. {\bf 8} 11 (2005), \texttt{arXiv:gr-qc/0601085}.

\bibitem{Bojowald1} M.~Bojowald, \emph{Quantum nature of cosmological bounce}, Gen.~Rel.~Gravit. {\bf 40} 2659--2683 (2008),
\texttt{arXiv:0801.4001 [gr-qc]}.

\bibitem{Rovelli} C.~Rovelli, \emph{Black hole entropy from loop quantum gravity}, Phys. Rev. Lett. {\bf 77} 3288--3291 (1996), \texttt{arXiv:gr-qc/9603063}.

\bibitem{ABC} A.~Ashtekar, J.~Baez, A.~Corichi and K.~Krasnov, \emph{Quantum geometry and black hole entropy}, Phys.~Rev.~Lett.~
{\bf 80} 904--907 (1998), \texttt{arXiv:gr-qc/9710007}.

\bibitem{Domagala} M.~Domagala and J.~Lewandowski, \emph{Black-hole entropy from quantum geometry}, Class.~Quantum~Grav. {\bf 21} 5233--5243 (2004), \texttt{arXiv:gr-qc/0407051}.

\bibitem{Agullo}I.~Agullo and P.~Singh, \emph{Loop Quantum Cosmology: A brief review}, in 100 years of General Relativity Vol. 4, Loop Quantum Gravity: The first 30 years, Editors A.~Ashtekar and J.~Pullin (World Scientific, 2017), \texttt{arXiv:1612.01236 [gr-qc]}.

\bibitem{Ashtekar} A.~Ashtekar and P.~Singh, \emph{Loop Quantum Cosmology: A Status Report}, Class. Quantum Grav.~{\bf 28} 213001 (2011), \texttt{arXiv:1108.0893 [gr-qc]}.

\bibitem{BF1}F.~Bayen, M.~Flato, C.~Fronsdal, A.~Lichnerowicz and D.~Sternheimer, \emph{Deformation theory and quantization I. Deformations of symplectic structures}, Ann.~Phys. {\bf 111} 61--110 (1978).

\bibitem{BF2}F.~Bayen, M.~Flato, C.~Fronsdal, A.~Lichnerowicz and D.~Sternheimer, \emph{Deformation theory and quantization. II. Physical applications}, Ann.~Phys. {\bf 111} 111--151 (1978).

\bibitem{Blaszak}M.~Blaszak and Z.~Doma\'{n}ski, \emph{Phase space quantum mechanics}, Ann.~Phys. {\bf 327}  167--211 (2012), \texttt{arXiv:1009.0150v2 [math-ph]}.

\bibitem{Curtright}T.~L.~Curtright, D.~B.~Fairlie and C.~K.~Zachos, \emph{A Concise Treatise On Quantum Mechanics In Phase Space} (World Scientific Publishing, 2014). 

\bibitem{Sahlmann}C.~J.~Fewster and H.~Sahlmann, \emph{Phase space quantization and Loop Quantum Cosmology: A Wigner function for the Bohr-compactified real line}, Class.~Quantum Grav.~{\bf 25} 225015 (2008), \texttt{arXiv:0804.2541v1 [math-ph]}.

\bibitem{Perlov}L.~Perlov, \emph{Uncertainty principle in loop quantum cosmology by Moyal formalism}, J.~Math.~Phys. {\bf 59} 032304 (2018), \texttt{ arXiv:1610.06532v4 [gr-qc]}.

\bibitem{Kempf}A.~Kempf, G.~Mangano and R.~B.~Mann, \emph{Hilbert Space Representation of the Minimal Length Uncertainty Relation}, Phys.~Rev.~{\bf D52} 1108--1118 (1995), \texttt{arXiv:hep-th/9412167v3}.

\bibitem{Camelia} G.~Amelino-Camelia, \emph{Introduction to quantum gravity phenomenology}, in Planck Scale Effects in Astrophysics and Cosmology, Lecture Notes in Physics, Vol.~669, G.~Amelino-Camelia and J.~Kowalski-Glikman (Eds.) (Springer-Verlag, Berlin, 2005).

\bibitem{Takhtajan}L.~A.~Takhtajan, \emph{Quantum Mechanics for Mathematicians}, Graduate Studies in Mathematics Vol. 95 (American Mathematical Society, 2008).

\bibitem{Strocchi}F.~Strocchi, \emph{An Introduction to the Mathematical Structure of Quantum Mechanics: A Short Course for Mathematicians}, Advanced Series in Mathematical Physics Vol.28 (World Scientific, 2008).

\bibitem{Folland}G.~B.~Folland, \emph{Harmonic Analysis in Phase Space} (Princeton University Press, 1989).

\bibitem{ReedI}M.~Reed and B.~Simon, \emph{Methods of Modern Mathematical Physics}, Vol.~I: Functional Analysis, 2nd ed. (Academic, San Diego, 1980). 

\bibitem{Compean}H.~Garc\'ia-Compean, J.~F.~Plebanski, M.~Przanowski, F.~J.~Turrubiates, \emph{Deformation Quantization of Classical Fields}, Int.~J.~Mod.~Phys. {\bf A16}, 2533--2558 (2001), \texttt{arXiv:hep-th/9909206v3}.



\bibitem{Kanwal}R.~P.~Kanwal, \emph{Generalized functions: Theory and Applications}, 3rd ed. (Springer Science, 2004).

\bibitem{Propagators}E.~Flores-Gonzalez, H.~A.~Morales-Tecotl and J.~D.~Reyes, \emph{Propagators in Polymer Quantum Mechanics}, Ann.~Phys. {\bf 336} 334--412 (2013), \texttt{arXiv:1302.1906v1 [math-ph]}.

\bibitem{Robustness}A.~Ashtekar, A.~Corichi and P.~Singh, \emph{Robustness of key features of loop quantum cosmology}, Phys.~Rev.~D {\bf 77} 024046 (2008), \texttt{arXiv:0710.3565 [gr-qc]}.

\bibitem{Rudin}W.~Rudin, \emph{Fourier Analysis on Groups}, Interscience Tracts
in Pure and Applied Mathematics (John Wiley and Sons, New York, 1962).

\bibitem{Hormann}G.~J.~H\"{ormann}, \emph{Limits of Regularizations for Generalized Function Solutions to the Schr\"{o}dinger Equation with ``Square Root of Delta" Initial Value}, Fourier Anal.~Appl. 1069--5869 (2017), \texttt{arXiv:1612.09055v2 [math.FA]}.

\bibitem{Blum}J.~Blum and B.~Eisenberg, \emph{Generalized summing sequences and the mean ergodic theorem}, Proc. Am. Math. Soc. {\bf 42} 423--429 (1974).

\bibitem{Teaching}A.~C.~Hirshfeld and P.~Henselder, \emph{Deformation quantization in the teaching of quantum mechanics}, Am.~J.~Phys. {\bf 70} 537--547 (2002), \texttt{arXiv:quant-ph/0208163v1}.

\bibitem{physicalCorichi}A.~Corichi, T.~Vukasinac and J.~A.~Zapata, \emph{Hamiltonian and physical Hilbert space in polymer quantum mechanics}, Class.~Quantum~Grav.~{\bf 24} 1495 (2007), \texttt{arXiv:gr-qc/0610072}.


\bibitem{Nozari}K.~Nozari and A.~Etemadi, \emph{Minimal length, maximal momentum, and Hilbert space representation of quantum mechanics}, Phys.~Rev.~D~{\bf 85} 104029 (2012), \texttt{arXiv:1205.0158 [hep-th]}.

\bibitem{Husain}G.~M.~Hossain, V.~Husain and S.~S.~Seahra, \emph{Background independent quantization and the uncertainty principle}, Class.~Quantum. Grav.~{\bf27} 165013 (2010), \texttt{arXiv:1003.2207 [gr-qc]}.

\bibitem{kho}
M.~Khodadi, 
K.~Nozari, 
S.~Dey, 
A.~Bhat and
M.~Faizal, 
\emph{A new bound on polymer quantization via an opto-mechanical setup},
Sci.~Rep. {\bf 8}  
1659 
(2018),
\texttt{arXiv:1801.00273 [gr-qc]}.


\bibitem{BojowaldLQC}M.~Bojowald, \emph{Loop Quantum Cosmology}, Living Rev.~Relativ.~{\bf 8} 11 (2005), \texttt{arXiv:gr-qc/0601085}. 

\bibitem{Koslowski}T.~A.~Koslowski, \emph{A Cosmological Sector in Loop Quantum Gravity}, \texttt{ 	arXiv:0711.1098 [gr-qc]}.

\bibitem{Marugan}G.~A.~Mena-Marugan, J.~Olmedo and T.~Paw\l{}owski, \emph{Prescriptions in loop quantum cosmology: A comparative analysis}, Phys.~Rev.~D~{\bf 84} 064012 (2011), \texttt{arXiv:1108.0829 [gr-qc].
}


\end{thebibliography}

\end{document}